# Different W cluster deposition regimes in pulsed laser ablation observed by *in situ* Scanning Tunneling Microscopy

D. Cattaneo, S. Foglio, C.S. Casari, A. Li Bassi, M. Passoni, C.E. Bottani

NEMAS-Center for NanoEngineered MAterials and Surfaces
CNISM-Dipartimento di Ingegneria Nucleare, Politecnico di Milano
Via Ponzio 34/3 I-20133 Milano, Italy

We report on how different cluster deposition regimes can be obtained and observed by *in situ* Scanning Tunneling Microscopy (STM) by exploiting deposition parameters in a pulsed laser deposition (PLD) process. Tungsten clusters were produced by nanosecond Pulsed Laser Ablation in Ar atmosphere at different pressures and deposited on Au(111) and HOPG surfaces. Deposition regimes including cluster deposition-diffusion-aggregation (DDA), cluster melting and coalescence and cluster implantation were observed, depending on background gas pressure and target-to-substrate distance which influence the kinetic energy of the ablated species. These parameters can thus be easily employed for surface modification by cluster bombardment, deposition of supported clusters and growth of films with different morphologies. The variation in cluster mobility on different substrates and its influence on aggregation and growth mechanisms has also been investigated.

## 1. Introduction

Great efforts have been devoted in recent times to the study of free and supported clusters [1-3] and to the deposition and aggregation mechanisms of clusters on different surfaces [4, 5], both to answer fundamental issues regarding properties of matter at the nanoscale, and to develop nanostructured films or surfaces with new and improved properties or tailored functions [2, 5-7]. In this context the comprehension of cluster formation and deposition and of the first stages of film formation (as a function of cluster size and energy) plays a fundamental role for the control of the film and surface properties. A variety of cluster deposition regimes have been extensively studied by means of theoretical models and experiments, mainly investigating the role of cluster deposition kinetic energy and, to some extent, of cluster size distribution [4, 5, 8-21]. It has been demonstrated that high energy deposition (i.e. several eV/atom or even more) leads to cluster penetration or cluster pinning [17-19]. In this way very smooth and compact films can be grown with an enhanced substrate adhesion [22]. On the other hand, low energy deposition (i.e. fractions of eV/atom) has been observed to lead to cluster diffusion and aggregation on the surface [4, 15], followed by juxtaposition or coalescence in larger clusters. This deposition regime has been exploited as a route for the synthesis of cluster-assembled materials retaining a *memory effect* of the precursor building unit properties [5].

Different cluster sources have been employed to obtain different deposition regimes [15, 18, 22, 23] and usually high kinetic energies are achieved by accelerating ionized clusters only, which results in a low deposition rate . Among various deposition techniques, Pulsed Laser Deposition (PLD) has been proposed as a versatile approach for the synthesis of both supported clusters and nanostructured films [24-28]. In fact, laser ablation in the presence of a background gas results in cluster aggregation and kinetic energy reduction during the plume expansion due to increased collision rate and cooling of the ablated species [27, 29-33], even though the dependence of cluster size and energy distribution as a function of gas pressure and distance from the ablation target is not well understood yet. An estimate of the kinetic energy of the deposited species has been reported in a few cases [27, 33-35]. In particular Irissou et al. have demonstrated the capability to tune the kinetic energy of Au species from several tens down to fractions of eV/atom by playing with gas pressure and target-to-substrate distance [27, 33]; nevertheless no compelling evidence of PLD cluster fragmentation or soft landing on impact, as expected from the above cited theoretical and experimental findings at the same kinetic energies, has been reported yet.

Of course the study of supported clusters greatly benefit from direct imaging and high resolution characterization (not only morphological) of clusters at surfaces. *Ex situ* Transmission Electron Microscopy (TEM), Atomic Force Microscopy (AFM) and Scanning Tunneling Microscopy (STM) have been widely employed to image clusters and nanoparticles deposited on different surfaces. Among other techniques, STM plays a peculiar role, because of its unique capability of providing high resolution direct imaging combined with spectroscopic information on a sub-nanometric scale. A great deal of work has been published concerning evaporation or atom-by-atom growth at the surface of supported clusters investigated by *in situ* STM; on the contrary, only a few *in situ* STM studies of cluster beam deposition on surfaces have been reported [23, 36-44], and to the best of our knowledge no *in situ* STM investigation of pulsed laser deposited clusters has been reported.

We here show, by means of *in situ* STM of W clusters deposited on atomically flat surfaces, namely HOPG and Au(111)/mica, that different deposition regimes can be achieved in pulsed laser deposition, when cluster size distribution, kinetic energy and interaction with the substrate are varied through a control of the deposition process in a background atmosphere and the use of substrates with different physical properties. In particular, cluster deposition and diffusion followed by juxtaposition, cluster melting at impact and subsequent diffusion and coalescence, cluster pinning and implantation can be obtained by varying PLD parameters such as the gas pressure and the substrate-to-target distance. The observation of different deposition regimes in the energy range of PLD is favored by the high cohesive energy (8.9 eV/atom) and high melting temperature (3695 K) of W, which is expected to lead to reduced fragmentation of clusters at low impact energies and to cluster pinning or crater formation at higher energies. Moreover, our observations reveal the important role of the substrate in the investigation of nanoparticles produced by PLD in the plasma plume since cluster mobility on the surface can be controlled by a careful choice of the substrate material.

## 2. Experiment

Clusters were synthesized by ablation of a W target with KrF laser pulses (pulse duration 10-15 ns, wavelength 248 nm, energy density ~ 4 J/cm$^2$) in an Ar background atmosphere and deposited at room temperature on HOPG and Au(111)/mica surfaces (details of preparation can be found in ref.[45]). Depositions at 10 Pa and 40 Pa Ar were performed at different substrate-to-target distances ($d_{ts}$ = 60 mm at 10 Pa; $d_{ts}$ = 30 mm at 40 Pa) and correspond to different kinetic energy regimes as explained in the following. In order to obtain isolated clusters, or cluster islands, a small number (1-30) of laser pulses was employed; the number of pulses was chosen to maintain an amount of deposited material corresponding to less than one equivalent monolayer. A Omicron UHV VT-SPM (base pressure 5·10$^{-9}$ Pa) is directly connected to the UHV-compatible PLD chamber (base pressure ~10$^{-7}$ Pa) via a magnetically coupled transfer system, allowing *in situ* characterization of the samples. HOPG substrates were prepared by cleavage in air followed by annealing at 200 °C for several hours in UHV; Au(111) surface was prepared in UHV by 15 min Ar$^+$ sputtering at 1 keV with the sample temperature at 550 °C . Such substrates maintain atomically flat surfaces when exposed to Ar background gas pressure in the deposition chamber (up to 100 Pa), as checked by STM measurements of the as prepared surfaces after gas exposure. STM constant current measurements were performed using home-made etched W tips. Typical STM tunneling currents (~ nA) often resulted in a drift of the observed deposited W aggregates, therefore images were acquired with a very low tunneling current (a few pA) and a relatively high tunneling potential (3-8 V) to increase the tip-to-sample distance and to correspondingly reduce the tip interaction with deposited clusters.

## 3. Results and Discussion

STM images of W clusters produced at 10 Pa Ar and 40 Pa Ar background gas pressure and deposited on HOPG are shown in fig. 1-A and 1-B. Particles with a regular shape (mean size 3.3 ±2 nm) are observed at 10 Pa Ar, while ramified islands are observed at 40 Pa Ar. These ramified structures look formed by the juxtaposition of nano-units of about 2-3 nm mean size. We believe that the larger particles observed on the surface at the lower background gas pressure (10 Pa, fig.1-A) are indeed the result of coalescence on the HOPG surface of smaller clusters formed in the ablation plume, as explained below.

This picture is supported by STM images of W clusters deposited in the same conditions but on a different substrate (fig. 1-C and 1-D). Deposition at 10 Pa Ar on Au(111) reveals the presence of holes, or craters (fig.1-C), suggesting a rather high kinetic energy of impinging species which have experienced a weak stopping action by the buffer gas. Conversely, deposits at 40 Pa Ar on Au(111) are characterized by isolated clusters and no craters (fig. 1-D), thus revealing a substantial lowering of the kinetic energy of the species reaching the substrate. Dolbec et al. [34] found kinetic energy values in the 4-45 eV/atom regime for the ablation of Pt deposited on HOPG (background He pressure of 65 Pa, laser fluence 4 J/cm$^2$), and with target-to-substrate distance in the range $d_{ts}$ = 30-60 mm. In addition Irissou et al. found kinetic energy values in the 0.1 – 50 eV/atom range for Au ablated in Ar at 4 J/cm$^2$ [33]. In particular they reported isoenergetic curves as a function of both Ar pressure and target-to-substrate distance (fig.4C in ref [33]). On the basis of these data (even though referred to different metals) we can estimate at least the order of magnitude of the kinetic energy of our W species to be in the range from some eV/atom for deposition at 10 Pa Ar and $d_{ts}$ = 60 mm (higher kinetic energy regime) to fractions of eV/atom, for deposition at 40 Pa Ar and $d_{ts}$ = 30 mm (lower kinetic energy regime). It has to be noticed that this estimate is only qualitative since the ablated elements are different; however the reported values for Pt and Au ([33, 34]) ablated with the same fluence (i.e. 4 J/cm$^2$, as in our case) show a similar behavior as a function of pressure and target-to-substrate distance. Moreover the measured kinetic energies are obtained from an emission line of the plume species (atoms, clusters, etc.) and are only representative of the cluster deposition kinetic energies. In any case this energy range is indeed compatible with the observed different cluster behavior on surfaces ranging from implantation to mobility and aggregation, also depending on the substrate surface properties.

From the observation of craters on Au(111) we have a signature of the impact of clusters (or even atoms) and the crater size is related to the size of the impinging species. Thus, craters provide information which is not related to aggregation phenomena occurring at the surface and influenced by mobility. Craters (fig.1-C, 2) may be ascribed to a complete penetration into the substrate, or to complete disintegration, although no residual particulate (only a rim at the crater edge) has been observed on the Au surface. They are not formed when high energy deposition is performed in the same conditions (i.e. 10 Pa and $d_{ts}$ = 60 mm) on HOPG (fig. 1-A, 2), probably due to the larger cohesion energy of graphite, which prevents implantation phenomena at this energy (bulk cohesive energy of W, Au and graphite are 8.9 eV/atom, 3.81 eV/atom and 7.37 eV/atom respectively [46], even though in principle surface bond energies should be compared). Carrol et al. [17] estimated the cluster kinetic energy, expressed in eV/atom, necessary for penetration of Ag cluster on HOPG to be of the order of the binding energy per C atom of the HOPG surface, and thus a lower value (a few eV/atom) can be expected for penetration in the Au surface, in agreement with the above predicted kinetic energy range. The discrepancy between crater size on Au(111) (average size below 1 nm, i.e. clusters composed by a few W atoms) and aggregate size on HOPG (average size 3.3 ±2 nm) can be better observed in fig. 2, where also the rim created by recoil after cluster impact can be observed. This proves that some aggregation process has occurred between species impinging on the HOPG surface, while the same species penetrate into the Au surface. Moreover on HOPG most of the aggregates have a nearly spherical shape, with a mean aspect ratio between diameter and height of about 1.5. We believe that the shape of the observed aggregates may be related to a resolidification of liquid droplets formed by coalescence of diffused clusters and even atoms, forming a larger cluster which does not retain any memory of the original building units. It has been proposed that very energetic species impinging onto the surface, if not able to properly transfer their energy to the substrate, can even melt [5] and, after diffusion, coalesce. In fact molecular dynamics simulations [5, 8, 9, 12] show that highly energetic cluster impacts generate a pressure shock wave and a transfer of the cluster kinetic energy to the substrate, followed by a very high local temperature increase. The coalescence between 'liquid' clusters is a process requiring energy, since it involves dislocation phenomena, elastic stress and formation of new chemical bonds as shown by Zhu and Averback for Cu clusters [47]. The resulting larger clusters then cool and become 'solid' in the most energetically favorable configuration, i.e. spherical. In summary, ablation at 10 Pa results in deposition of species with a high kinetic energy leading to penetration in the Au(111) surface, and to diffusion/coalescence on the HOPG surface. Substrate properties such as surface cohesive energy and diffusivity may in some cases modify or even hide effects related with cluster impact. This may explain for instance why Dolbec et al. [34] observe that the mean size of Pt particles on HOPG mainly depends on the coverage (i.e. number of laser pulses) while apparently not depending on the kinetic energy (i.e. background gas pressure and target-to-substrate distance).

Depositions at a lower kinetic energy (40 Pa Ar, $d_{ts}$ = 30 mm) show no craters on both Au(111) and HOPG (figs. 1-B, 1-D, 3), indicating a less energetic impact with respect to deposition at 10 Pa. It is likely that the mobility of W species on Au(111) is smaller than on HOPG [21], and that most of the aggregates observed on the Au surface are formed in the plume before impact on the substrate, not being (at least most of them) the result of a post-deposition surface aggregation. This is evidenced by deposition on HOPG, where irregular, ramified aggregates are observed, composed by juxtaposed substructures having nearly the same size as clusters observed on Au(111), as confirmed by a statistical analysis of their size distribution. High resolution line profiles in fig.3 reveal that clusters arriving at the HOPG and Au(111) surface (and formed in the ablation plume in the same conditions) have the same size, corresponding to clusters composed by a few hundreds atoms. When deposition is performed at 40 Pa on HOPG, at difference with deposits at 10 Pa, clusters only stick to other clusters probably because they overcome a critical size above which coalescence is no longer favored and/or they do not possess enough energy to remain in the liquid phase [5]; probably both a smaller kinetic energy and a larger average cluster size lead to the described process. This kind of growth has been largely studied in the literature both experimentally and with theoretical models (DDA, Deposition-Diffusion-Aggregation [4, 14, 15]). Therefore, differences between structures observed on Au(111) and HOPG in the same deposition conditions (i.e. 40 Pa) are a consequence of the different mobility of W species on these substrates. Particles observed at 40 Pa on Au(111) are clusters formed in the ablation plume which impinge onto the Au surface and experience a negligible diffusion, maybe also because some of them may be partially implanted. In fact, the cluster aspect ratio (between diameter and height) is about 2, i.e. clusters are flattened, suggesting a not negligible impact energy, or at least a complex cluster-substrate interaction. This observation opens the way to the study of the size distribution of clusters resulting from ablation in different conditions (e.g. different gas pressure) by observation of deposits on a flat substrate where the mobility is strongly reduced. The reported behavior shows the fundamental role of the substrate when investigating species formed in the ablation plume, i.e. observed isolated structures on surfaces are deeply influenced by interaction with the substrate itself. Anyway, this effect should vanish in film growth with increasing deposited material and above a certain film thickness. We have already reported on tungsten thin films (about one hundred nm thick) [48], showing different morphology from compact and smooth to nanostructured and porous moving from a few Pa to several tens Pa Ar deposition pressure, in agreement with the cluster deposition regimes here observed. A similar trend was also observed for different metals [49] and semiconductors [35, 50] supporting the relevance of our results for a wide range of materials.

## 4. Conclusions

In summary, we have shown that by means of *in situ* STM measurements of W clusters produced by laser ablation and deposited on proper surfaces it is possible to observe a variety of cluster deposition and aggregation mechanisms, ranging from cluster implantation to soft landing. This permits to really unveil the behavior on surfaces of PLD clusters. By exploiting simple PLD deposition parameters, such as the background gas pressure and the target-to-substrate distance, it is in principle possible to vary both the size and the kinetic energy of the deposited clusters, thus making it possible to synthesize and deposit clusters at very different deposition regimes. Therefore laser ablated clusters can be exploited for either the deposition of smooth films by cluster fragmentation and energetic deposition or as building blocks for the synthesis of nanostructured films by cluster soft landing and assembling [48]. We clearly show that PLD may be used as a flexible cluster source taking advantage of the extremely high versatility of the technique (e.g. wide variety of target materials even with complex stoichiometry). Also, the understanding of thin film growth mechanisms may in principle open the possibility of investigating how structural properties of functional materials can be tuned, and how the parameters influencing the laser ablation process can be employed as controlling parameters of the thin film structure in the deposition process of functional coatings. Our results demonstrate that the combination of a PLD system with a directly coupled STM provides a unique experimental system for the *in situ* investigation of morphological properties of supported clusters and thin films. Moreover, other properties (e.g. electronic, catalytic, etc.) and phenomena (e.g. oxidation, or other chemical reactions occurring at the surface) can be investigated by combining STM and scanning tunneling spectroscopy (STS) measurements, resulting in a powerful tool for the study of fundamental properties of model surface nanostructures at the atomic scale.

**Figure captions**

Figure 1
STM images (100x100 nm$^2$) of W deposits on HOPG (A): 10 Pa, (B): 40 Pa and on Au(111): (C) 10 Pa, (D) 40 Pa.

Figure 2
STM images (15x15 nm$^2$) and line profiles of W deposits at 10 Pa Ar, $d_{ts}$ = 60 mm (higher deposition energy regime, see text) on Au(111) (left) and HOPG (right).

Figure 3
STM images (40x40 nm$^2$) and line profiles of W deposits at 40 Pa Ar, $d_{ts}$ = 30 mm (lower deposition energy regime, see text) on Au(111) (left) and HOPG (right).

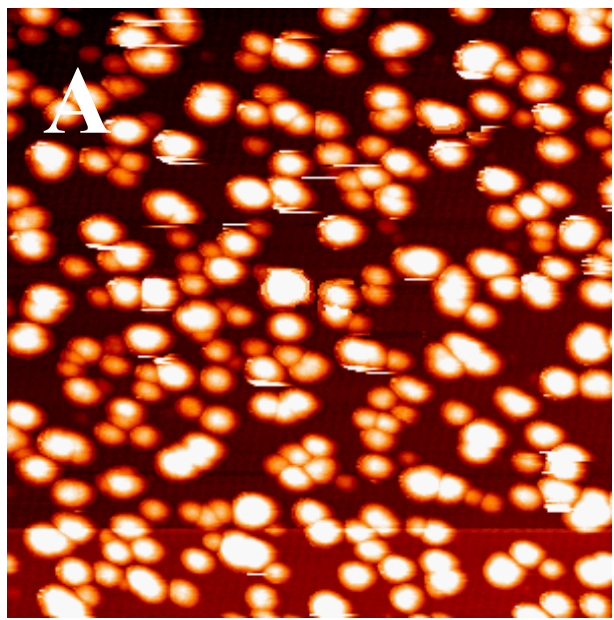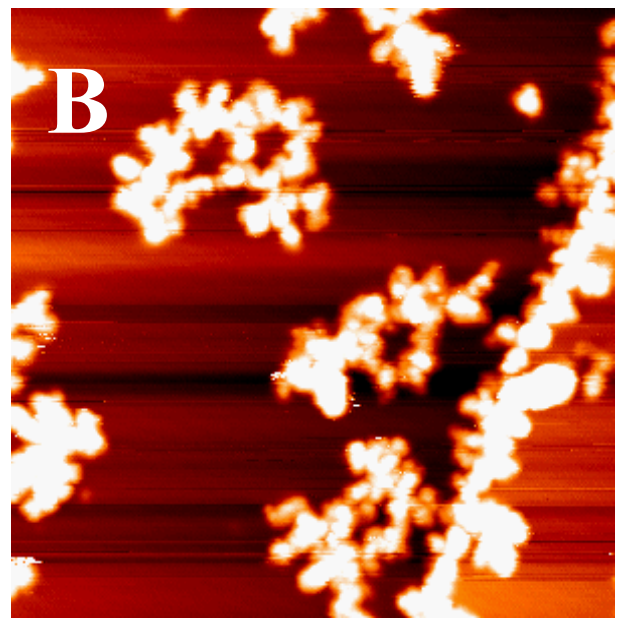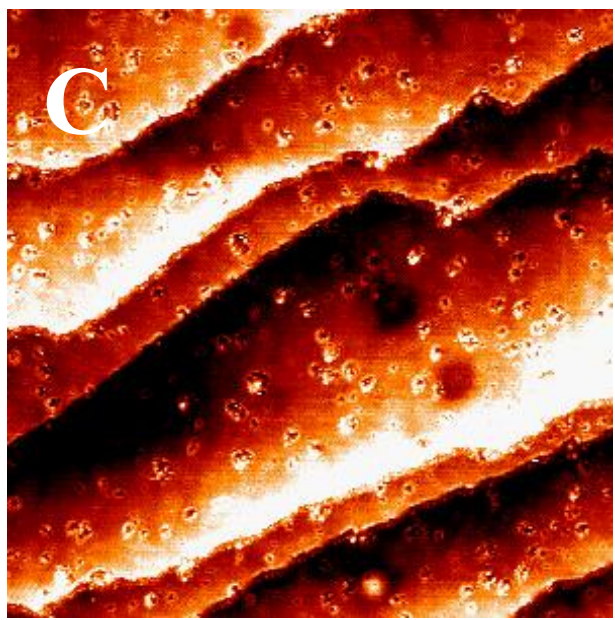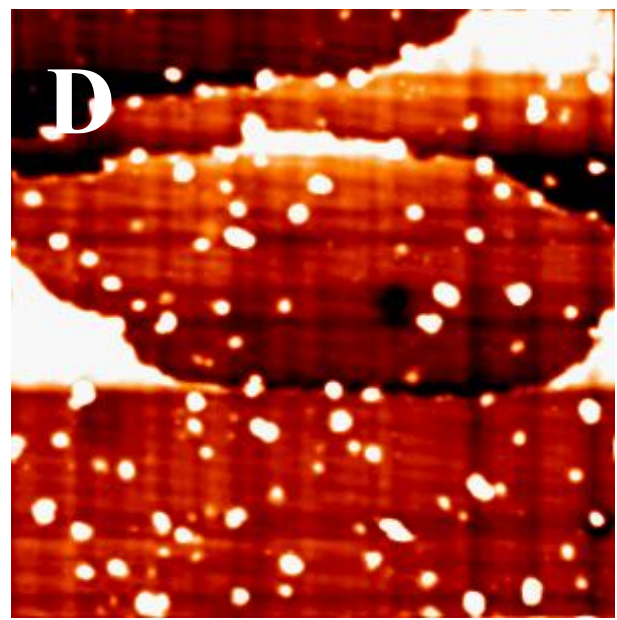

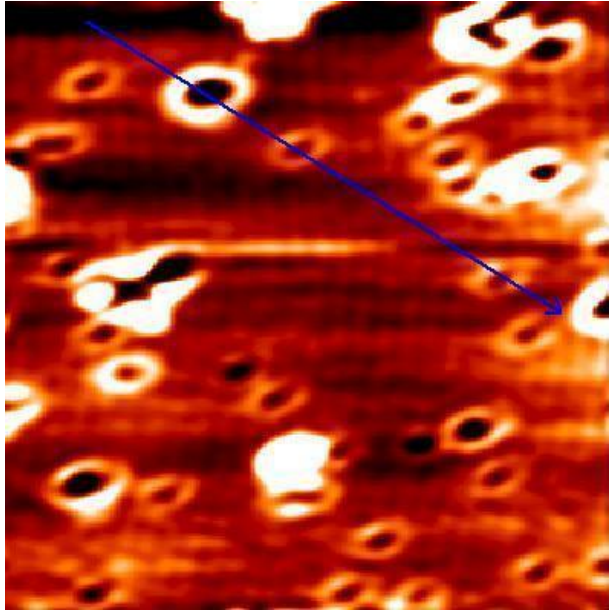
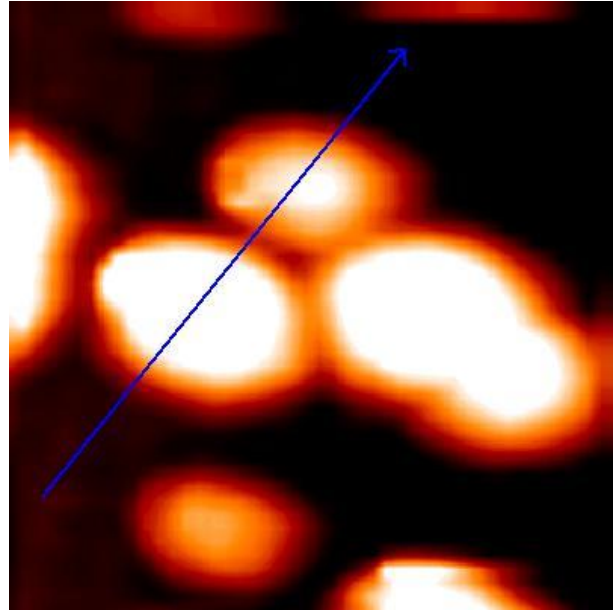
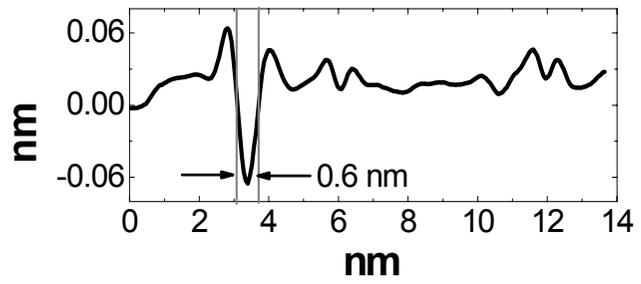
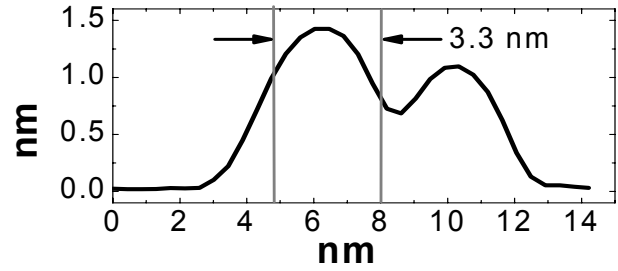

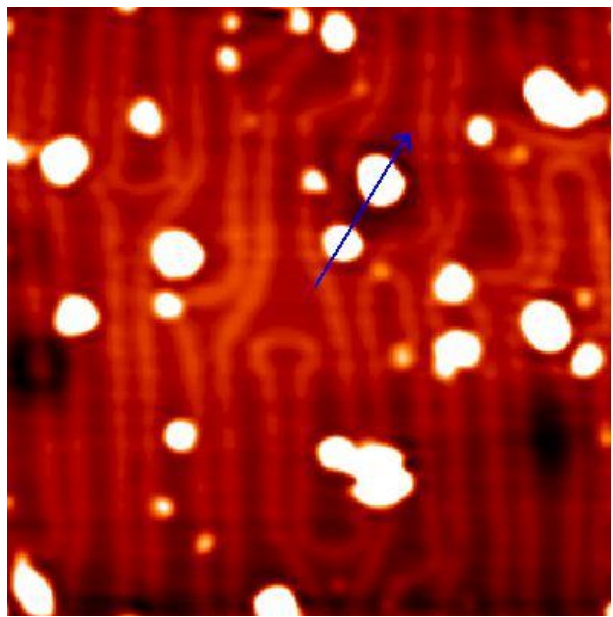
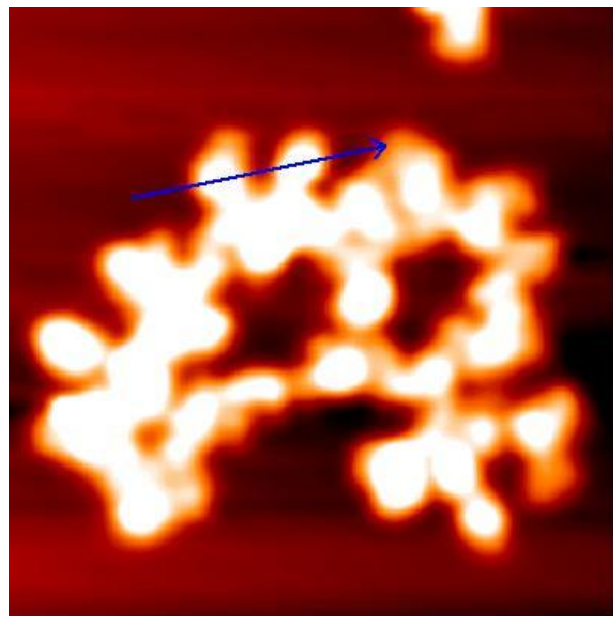
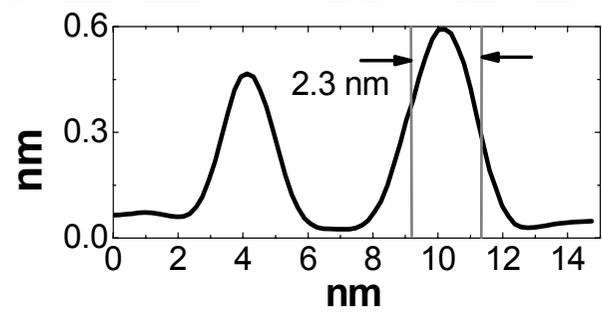
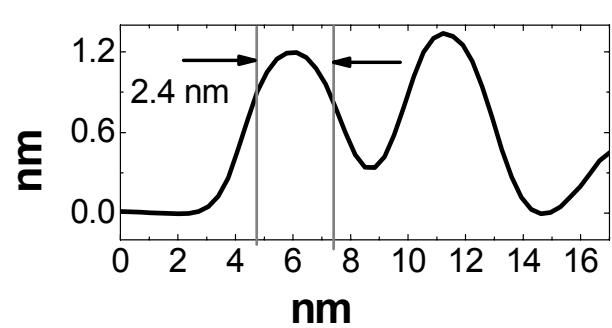